# Anomalous diameter dependence of thermal transport in ultra-narrow Si nanowires

Hossein Karamitaheri[1,2], Neophytos Neophytou[1,3], and Hans Kosina[1]

[1]Institute for Microelectronics, TU Wien, Gußhausstraße 27-29/E360, A-1040 Wien, Austria
[2]Department of Electrical Engineering, University of Kashan, Kashan, 87317-51167, Iran
[3]School of Engineering, University of Warwick, Coventry, CV4 7AL, UK
E-mail: {karami | neophytou | kosina}@iue.tuwien.ac.at

## Abstract

We present atomistic valence force field calculations of thermal transport in Si nanowires of diameters from 12nm down to 1nm. We show that as the diameter is reduced, the phonon density-of-states and transmission function acquire a finite value at low frequency, in contrast to approaching zero as in the bulk material. It turns out that this effect results in what Ziman described as the *"problem of long longitudinal waves"* [1], which states that the thermal conductivity of a material increases as its length is increased due to the vanishing scattering for long-wavelength phonons. We show that this thermal transport improvement also appears in nanowires as their diameter is decreased below *D*=5nm (not only as the length increases) originating from the increase in the density of the long wavevector modes. The observation is present under ballistic transport conditions, and further enhanced with the introduction of phonon-phonon scattering. Because of this, in such ultra-narrow nanowires, as the diameter is reduced, phonon transport is dominated more and more by lower energy phonons with longer mean-free paths. We show that ~80% of the heat is carried by phonons with energies less than 5meV, most with mean-free paths of several hundreds of nanometers.

**Keywords:** Confined phonons, silicon nanowires, thermal transport, long wavelength phonons.

# I. Introduction

In bulk semiconductors and insulators the thermal resistance arises from phonon-phonon scattering because of the anharmonicity of the inter-atomic potential. At room temperature, phonon-phonon scattering processes are strong and dominate the behavior of the thermal conductivity ($\kappa_l$). A large part of the heat in semiconductors is carried by long wavelength longitudinal phonons, which have extremely long mean-free-paths (MFP) for scattering as well. In fact, it was pointed out that the MFP of the long wavelength longitudinal phonons diverges as their frequency tends to zero, resulting in the thermal conductivity to be a function of the size of the bulk solid, and diverging as the size of the solid increases [1, 2]. The most commonly employed single-mode-relaxation-time (SMRT) approximation for the solution of the Boltzmann transport equation (BTE) for phonons [3, 4], uses an $\omega^2$-dependent phonon-phonon (Umklapp) scattering rate [5, 6]. This model, in combination with the 3D bulk density-of-states, which is proportional to the $\omega^2$ at low frequencies, removes this ambiguity, and successfully explains the thermal conductivity of various semiconductors over a wide range of temperatures (also after appropriately including other common scattering mechanisms, such as defect scattering and grain-boundary scattering).

The thermal conductivity of 1D channels such as carbon nanotubes (CNTs), nanoribbons, and silicon nanowires (NWs) has also been addressed in several recent studies [7, 8, 9, 10, 11, 12, 13, 14, 15, 16], since such channels attracted significant attention for heat management and thermoelectric applications [17, 18, 19]. The divergence of $\kappa_l$ with the size of the solid, or the *"problem of long longitudinal waves"* as referred to by Ziman [1], is stronger in this case, since the density-of-states (*DOS*) in 1D structures is no longer $\omega^2$-dependent, but has a finite value even at $\omega = 0$ [20], which increases the importance of low wavevector phonons. Indeed, several theoretical and experimental works have pointed out that thermal conductivity in 1D systems deviates from Fourier's law, or even increases with increasing channel length, either linearly, logarithmically, or following some power law [7, 21, 8, 9, 22, 23]. By including additional scattering mechanisms to the 3-phonon Umklapp mechanism usually employed, such as terms constant or linear in *ω* [16, 24, 25], 3-phonon processes of second order [7], highly anharmonic potentials [26, 27], employing the exact solution of the Boltzmann equation [21], or molecular dynamics (MD) [10], the divergence is reduced, but it is still persistent, especially at low temperatures.

In this work, we revisit this problem for ultra-thin Si NWs of diameters below 12nm using the atomistic modified-valence-force-field (MVFF) method for the calculation of the phonon modes and two approaches for phonon transport: i) the Landauer approach for ballistic transport, and ii) the BTE for diffusive phonon transport. We show that the issue of long-wavelength phonons turns out to be much more significant in 1D systems compared to the bulk material: 1) not only for low temperatures, but also at room temperature, 2) and not only as the length of the channel is increased, but as the diameter is reduced as well, resulting in an improvement of thermal transport with increasing confinement. This was previously reported using MD simulations [10], however, in this work we explain it in simple terms related to: i) the increase in the *DOS* of the low frequency, low wavevector modes with diameter reduction, ii) the narrowing of the so-called "phonon window function" that determines phonon transport, favoring low wavevectors. We show that this effect does not qualitatively depend on the nature of scattering or the scattering model itself. It is observed under ballistic conditions, whereas the introduction of phonon-phonon or moderate phonon-boundary scattering mechanisms pronounce it even more, as long as high frequency modes are scattered more that low frequency modes (as for example in the case of the usual 3D bulk phonon scattering models that we also employ here). Our calculations indicate that more than 80% of the heat is carried by low energy, long wavelength phonons.

## II. Approach

**<u>The problem of long longitudinal waves</u>**: When considering the classical description of the energy distribution of phonons, in the Debye approximation, the heat conductivity of acoustic modes is given by [1, 9]:

$$\kappa_l = \frac{k_B}{3\Omega} \int_{\omega_{min}}^{\omega_D} v_s^2 \tau(\omega) DOS(\omega) d\omega \tag{1}$$

where $v_s$ is the corresponding sound velocity, $\omega_{min} = 2\pi v_s / L$ is the minimum allowed phonon frequency which is determined by the longest allowed wavelength at a given channel length, and $\omega_D$ is the Debye frequency. As the bulk density-of-states is proportional to $DOS(\omega) : \omega^2$, the Umklapp-limited thermal conductivity in 3D (with usually $\tau_U : 1/\omega^2$ [5, 6]) is bounded, even when the contribution of the long MFP phonons (as $\omega \to 0$) is included:

$$\kappa_l \sim \int_{2\pi v_s/L}^{\omega_D} v_s^2 d\omega \sim \omega_D - 2\pi v_s / L \tag{2}$$

since the second term of Eq. 2 goes to zero as the length increases.

In the case of 1D structures, on the other hand, the $DOS$ is finite for low frequencies $DOS(\omega) = L/\pi v_s$ [28], which does not allow the cancellation of the $\tau(\omega) \sim 1/\omega^2$ term. Therefore, the Umklapp-limited thermal conductivity of a 1D system is given by:

$$\kappa_l \propto \int_{2\pi v_s/L}^{\omega_D} \frac{1}{\omega^2} d\omega \sim L \tag{3}$$

indicating a linear divergence with the channel length. Physically, this means that the lowest wavevector that contributes to thermal conductivity is determined by the length of the channel. For an infinite channel, longer and longer wavevectors of finite $DOS$ are involved, which causes divergence in the thermal conductivity, in contrast to bulk. Several works in the literature attempt to add corrections that would bound $\kappa_l$ as the channel length increases [7, 21], although large $\kappa_l$ could still be possible in 1D. Some authors still use the bulk dispersion even for ultra-narrow channels [29, 30, 31], others include a constant specularity parameter for surface roughness, that adds a constant term in the total scattering rate, and removes the singularity for $\omega \to 0$ [12, 32, 33, 34]. A different approach was proposed recently by Mingo *et al.*, where an additional rate of a second order 3-phonon scattering mechanism without $\omega$-dependence was introduced for thermal transport in carbon nanotubes as [7]:

$$\frac{1}{\tau_{U2}} = A_0 T^2 \qquad (4)$$

where $A_0$ is a frequency independent constant. This is just an order of magnitude approximation which removes the singularity for low frequency phonons, although very high thermal conductivities are still achieved. A study of the exact solution of the phonon BTE in carbon nanotubes showed a saturation in $\kappa_l$ as the length increased to the millimeter range [21], which could prove the suitability of using the $\tau_{U2}$ of Eq. 4.

The importance of the long wavelength phonons, however, is not only pronounced by the length of the channel. In this work we show that the reduction in the diameter of the NW can also result in the same effect. This is attributed to two important events: i) NWs have a finite phonon *DOS* at low frequencies that also increases as the diameter is reduced, in contrast to bulk. ii) Low temperature or scattering events narrow the so-called "phonon window function" preferentially towards these low-frequency phonons, which then tend to dominate thermal transport. This is demonstrated under both, ballistic, and diffusive transport conditions.

**Low-dimensional phonon transmission and DOS:** For the calculation of the phononic bandstructure we employ the modified valence force field method (MVFF) [35, 36]. In this method the interatomic potential is modeled by the following bond deformations: bond-stretching, bond-bending, cross-bond-stretching, cross-bond-bending-stretching, and coplanar-bond-bending [35]. The model accurately captures the bulk Si phonon spectrum as well as the effects of confinement [36]. The model and its validity for nanoscale Si and Si NWs is presented in previous works [36, 14]. It has the advantage of being able to treat atomistically relatively large NW diameters, up to 12nm, with thousands of atoms in the cross section unit cell, and therefore can capture the transition of the phononic dispersion and of the thermal properties from 3D to 1D channels. Using the phononic dispersion, the density of states and the transmission (number of modes at given energy) are calculated as follows:

$$DOS(\omega) = \sum_i DOS_i(\omega) = \sum_i \sum_q \delta(\omega - \omega_i(q)), \qquad (5)$$

$$\overline{T}_{ph}(\omega) = M(\omega) = \frac{\pi}{2} \sum_{i,q} \delta(\omega - \omega_i(q)) \frac{\partial \omega_i(q)}{\partial q}. \qquad (6)$$

The transition from the 3D-like features to 1D-like features is shown in Fig. 1 which depicts the transmission function per unit area as a function of frequency for NWs of diameters $D$=12nm, 2nm, and 1nm. The inset of Fig. 1 shows the density of phonon states per unit volume of these NWs versus frequency. Note that we use the <100> transport orientation throughout this manuscript. The basic features we describe, however, are valid for different orientations as well. The transmission of the 12nm NW as well as the $DOS$ follow the usual $\omega^2$ relation at low frequencies. For the thinner diameters, however, the transmission and $DOS$ are constant/finite at low frequencies, and increase as the diameter is reduced. The fact that there is a constant transmission at $\omega$=0 rad/s for the acoustic branches leads to a strong manifestation of the *"long-wavelength problem"* in the thermal transport. The density, and importance, of the long-wavelength modes in carrying heat increases, which improves thermal transport as the NW diameter is reduced. Below we examine this effect under ballistic transport conditions and afterwards extend the findings to diffusive phonon transport conditions.

## II. Results

**Ballistic transport:** An effective way to demonstrate the influence of the increasing low frequency phonons on thermal transport, is by examining the differential (or frequency spectrum) of the ballistic thermal conductance versus energy. This quantity, (normalized by the NW area $A$) is calculated as [37, 38, 14]:

$$\frac{dK_l(\omega)}{Ad\omega} = \frac{\pi k_B^2 T}{6A} \sum_i \int v_{g,i}(q) W_{ph.}(\hbar\omega) \delta(\omega - \omega_i(q)) dq \qquad (7)$$

where $W_{ph.}(\hbar\omega)$ is the phonon window function that determines the conductance, defined as [39, 13]:

$$W_{ph.}(\hbar\omega) = \frac{3}{\pi^2 k_B T}\left[\frac{\hbar\omega}{k_B T}\right]^2 \frac{e^{\hbar\omega/k_B T}}{\left(e^{\hbar\omega/k_B T}-1\right)^2} \quad (8)$$

This function has strong temperature dependence as shown in Fig. 2. At 300K, the phonon window function (blue line in Fig. 2) is a wide and flat function, covering most of the energy spectrum [13]. Therefore, under ballistic conditions at room temperature, the entire energy spectrum contributes to the thermal conductance. For lower temperatures, on the other hand, the window function becomes narrower, and preferentially weighs the low frequency phonon contribution to thermal transport, whereas the high frequency phonons do not contribute.

The differential of the ballistic conductance for NWs of different diameters is shown in Fig. 3a for the low phonon energy region up to 10meV (the energy region we are mostly interested in). Here, room temperature, T=300K, is used. Clearly, the contribution of low frequency phonons increases as the NW diameter is decreased, a consequence in the increase in their density of states. Note that the contribution of high frequency phonons, however, decreases compared to NWs of larger diameters, as shown in the inset of Fig. 3a. This is because confinement reduces the number of phonon branches, primarily at higher frequencies. Figure 3b shows the same quantity but for lower temperature, T=20K. The contribution of low frequency phonons increases also in the case of lower temperatures.

The effect of low wavevector modes, however, is not noticeable on the total thermal conductance at room temperature. The total thermal conductance of the NWs versus diameter for various temperatures is shown in Fig. 4. At room temperature the ballistic thermal conductance of NWs is almost diameter invariant (black line for T=300K in Fig. 4). The reason is because the contribution of the low-frequency modes with diameter reduction is small compared to the contribution of the rest of the spectrum, which consists of a large number of subbands, and at room temperature, most of the spectrum contributes to transport due to the flat phonon window function. The situation is different at lower temperatures where the window function is much narrower (Fig. 2). In that case, only the low energy part of the phonon spectrum contributes to transport, which has a strong NW diameter dependence, as shown in Fig. 3b. The overall thermal conductance at lower temperatures in Fig. 4, therefore, increases as the diameter is reduced. An increase of ~5x is observed at T=20K as the diameter is decreased from $D$=5nm to $D$=1nm (green-triangle line). The increase is reduced as the temperature raises, and at 300K no benefit is observed in thermal transport with diameter reduction.

The important observation here is that the increase in the density of low-frequency, low wavevector phonon modes, becomes important in heat transport when the phonon window function becomes narrower. Low temperature is one way that causes window narrowing. Below we examine the influence of the increasing low wavevector phonon density under diffusing conditions and discuss whether a conductivity increase could also be observed.

**Diffusive transport:** In the case of diffusive transport, where phonons undergo phonon-phonon or phonon-boundary scattering, the window function is multiplied by the phonon-phonon scattering lifetime as $\tau_{ph.-ph.}W_{ph.}$. In general, the low-frequency, low-wavevector modes undergo weaker scattering, compared to the high frequency modes [5, 6, 39]. This also makes the window function narrower in energy, preferably selecting the low frequency modes for transport.

To investigate the influence of the low frequency modes in narrow NW diameters under diffusive thermal transport conditions, we employ the phonon lifetime approximation in the phononic Boltzmann transport equation as [3, 29]:

$$\kappa_l = k_B \sum_{i,q} \tau_i(q) v_{g,i}(q)^2 \left[\frac{\hbar\omega_i(q)}{k_B T}\right]^2 \frac{e^{\hbar\omega_i(q)/k_B T}}{\left(e^{\hbar\omega_i(q)/k_B T}-1\right)^2} \quad (9)$$

where $v_{g,i}(q) = \partial\omega_i(q)/\partial q$ is the group velocity of a phonon with wavevector $q$ in subband $i$, and $\tau_i(q)$ is the scattering time. For the calculation of the relaxation times, we still adopt the usual bulk formalism for Umklapp scattering employed in the literature for 3D phonons [5, 6]:

$$\frac{1}{\tau_U} = B\omega_i(q)^2 T \exp(-\frac{C}{T}) \qquad (10)$$

where $B = 2.8 \times 10^{-19}$ s/K and $C = 140$ K [39].

To the phonon-phonon scattering rate of Eq. 10, we added the scattering mechanism $A_0 T^2$ as proposed by Mingo *et* al. and described by Eq. 4 using Matthiessen's rule. We then calibrated our phonon-phonon thermal conductivity calculations for Si NWs against the MD simulation results of Donadio *et* al. [10] and the results of Luisier [40] for NW diameters up to $D$=5nm as shown in Fig. 5a. These calculations were performed assuming room temperature $T$=300K and NWs in the <100> transport direction. The parameter $A_0$ of the 3-phonon second order processes was set to $A_0$=15000/sK$^2$, which provides a good agreement between our results and these two other studies for the entire diameter range considered. Our results are also in good agreement with these studies for temperatures other than room temperature (not shown here). A clear increase in the thermal conductivity by ~5x is observed as the diameter is decreased. This increase can be directly attributed to the increasing importance of the low-frequency longitudinal modes as the diameter is reduced, after the transmission and *DOS* acquire a finite value (Fig. 1). Indications about thermal conductivity improvements due to phonon confinement can be found in other works as well, for Si NWs and other materials [8, 11, 10, 30, 41]. This is the first time, however, that we connect this increase to the finite value that the phonon *DOS* acquires for the low-frequency phonons.

We need to stress here, that we do not claim that the scattering model described by Eq. 10, and originally used and calibrated to 3D bulk phonons, is fully appropriate for the ultra-narrow nanowires considered in this study, whose modes we describe with 1D dispersions as well. However, although the model assumes 3D phonons, it is commonly employed in the calculation of the thermal conductivity in nanostructures [12, 42]. The reason why such a model derived for 3D channels provides sufficient accuracy for nanostructures, is that even for 1D channels the atomic vibrations are still in 3D and not constrained in one particular direction [7, 12]. And unlike the case of electronic transport, the phonon Umklapp scattering rate is not in general proportional to *DOS* which only in 3D follows a $\omega^2$ dependence [25]. The reasons are discussed by Hepplestone in Ref. [25] as follows: The three-phonon scattering is a consequence of the cubic anharmonic term of elastic potential $V_3$, which involves the derivatives of the potential with respect to displacement vectors in 3D. The displacement vector is related to frequency which appears eventually in relaxation time. As the vibration of atoms is always in 3D (even in low dimensional systems) the form of relaxation time in one-dimensional systems is expected to be close to that of 3D bulk materials.

There are additional explanations why the scattering rates will not increase as the *DOS* increases. 1D dispersions make it more difficult to satisfy simultaneously the momentum and energy conservation for the three phonon scattering. In 3D, for example, a surface in the Brillouin zone of available states exists that satisfies these requirements. In 2D the surface becomes a line, whereas in 1D only single points satisfy such requirements, which makes low frequency phonons less subjective to scattering [7]. This is the reason that even in more sophisticated MD calculations, the conductivity still diverges and the scattering rates do not follow the increase in the *DOS* of the low frequency phonons [10].

Modifications and extensions of the simplified 3D bulk $\omega^2$ transport model are described in the literature to improve its validity for nanostructures, still under the assumption of 3D phonons [43, 44]. The modification proposed by Mingo *et* al. [7], which includes an additional frequency independent scattering mechanism $\sim A_0T^2$, is the one we employ in this work. Other models, derived specifically for 1D channels i.e. case carbon nanotubes [16, 24], propose the addition of both a constant and a linear frequency term to the already squared-frequency term. The scattering rate is then proportional to $\sim A_0+(A_1T)\omega+(A_2T)\omega^2$, which is also dominated by the $\omega^2$ term [24, 16].

The main point here, however, is that in all cases any additional term introduces larger scattering for higher frequencies. The consequence of this is the narrowing of the diffusive phonon window function which is now multiplied by the relaxation times, $\tau_{ph.-ph.}W_{ph.}$. Figure 5b compares the ballistic window function to the diffusive window function, both at room temperature. The diffusive window function then has a similar narrowing shape as the one for lower temperatures in Fig. 2, and preferentially increases the importance of low-frequency phonons. The stronger scattering for high frequency phonons compared to low frequency ones, or equivalently longer mean-free-paths for low-frequency phonons, is a general feature, also supported by more sophisticated calculations for different nanostructures and different scattering mechanisms as well [45, 46, 47]. That explains why the traditional simplified Umklapp model based on 3D phonons still works adequately for nanostructures (at least qualitatively). In different nanochannel cases the actual physics of scattering could significantly vary, but if this feature is present, we claim that the density of low-frequency modes will increase with diameter reduction, and the thermal conductivity can increase as well. Indeed, our calculations in Fig. 5a show that for the $D$=1nm NW the thermal conductivity can increase back to the bulk value, in very good agreement with molecular dynamics calculations [10], although the accuracy of our model is questionable at such low diameters. We should also note that this narrowing of the $\tau_{ph.-ph.}W_{ph.}$ function is not sensitive to the value of $A_0$ we use to calibrate our data, even if $A_0$ increases or decreases by orders of magnitude.

## IV. Discussion

In this section, we discuss three consequences of the increase in the long wavevector, low frequency mode density, and their increasing importance in thermal transport through the phonon window narrowing. As the NW diameter is reduced: i) a larger part of the heat is carried by lower frequency phonons, ii) a larger part of the heat is carried by longer mean-free-path phonons, iii) the detrimental effect of phonon-boundary scattering could be partially compensated by the increase of the longitudinal mode density under moderate roughness amplitudes.

The importance of the longitudinal modes is indicated in the colorplot of Fig. 6a which shows the contribution of each phonon state to the ballistic thermal conductance of the $D$=2nm NW at T=300K. The dark-red color of the longitudinal acoustic (LA) mode indicates that most of the heat is carried by this low frequency, low wavevector mode. A large contribution is also attributed to the transverse acoustic (TA) and flexurial acoustic (FA) modes. The phonon states on the higher energy quasi-optical and optical modes carry only little heat due to their low phonon group velocities. To quantify the increasing importance of the low frequency modes with decreasing diameter, Fig. 6b shows the cumulative ballistic thermal conductance versus energy at room temperature. In bulk Si, the low frequency modes carry little heat under ballistic conditions as indicated by the black line from the work of Jeong *et* al. [39]. Nanowires with larger diameters, e.g. $D$=12nm, exhibit similar behavior, as shown by the green line marked by triangles. For narrower diameter NWs, lower energy phonons become even more important. For $D$=2nm and $D$=1nm, ~40% and ~60% of the heat, respectively, is carried by phonons with energies below 20meV. It is interesting to notice the steep slope of the $D$=1nm curve (blue line) at low energies indicating the increase in the mode density at low frequencies. The steepness abruptly decreases after $E = 60$meV, indicating the decrease in the mode density at higher energies.

The corresponding results for the case of phonon-phonon limited scattering conditions are shown in Fig. 6c and Fig. 6d. In this case, the low-frequency modes carry most of the heat even in the bulk material, as indicated by the black line from the work of Jeong *et* al. [39]. Almost half of the heat is carried by phonons of energies below 10meV. Nanowires with larger diameters, e.g. $D$=12nm, again exhibit similar behavior, as shown by the green line marked by triangles, and as under the ballistic conditions, for narrower NW diameters lower energy phonons become even more important. For $D$=2nm and $D$=1nm, ~80% of the heat is carried by phonons with energies below 5meV. The importance of the low-frequency modes, therefore, is stronger compared to the ballistic case, and the increasing density of these modes as the NW diameter is reduced could lead to most of the heat being carried by such low frequencies.

A useful quantity that more clearly demonstrates the importance of the differences between the thermal conductivity of the NWs with larger diameters versus the NWs with ultra-narrow diameters is the cumulative phonon-phonon scattering limited thermal conductivity versus the phonon MFP. The MFP for a phonon of momentum $q$ in branch $i$, is calculated as $\lambda_i(q) = \tau_i(q) v_i(q)$, where $v_i(q)$ is the phonon group velocity. This is shown in Fig. 7 for the NWs with diameters from $D=12$nm (green line) down to $D=1$nm (black line). For the wider NWs, the heat is carried almost uniformly by phonons of MFPs from one nanometer to several micrometers, as in the case of bulk Si [39, 48]. As the diameter is reduced, however, the portion of phonons with long MFPs increases. For the $D=2$nm and $D=1$nm NWs, for example, a large part of the thermal conductivity, almost 80%, is carried by phonons of MFPs longer than 1μm, corresponding to the low-frequency, long-wavelength phonons.

We finally note that in our work we mostly considered the effect of increasing low frequency mode density on the ballistic thermal conductance, and provided possible indication of its even stronger importance in the case of phonon-phonon limited scattering conditions. Ultra-narrow NWs, however, suffer from enhanced phonon-boundary scattering, which is the main reason the Si nanostructures are promising thermoelectric materials [49, 17]. We point out here that the increase in the thermal conductance because of the increase in low frequency modes could also be noticeable under moderate phonon-boundary scattering conditions as well. The reason is again that even in this case, the high frequency "flatter" quasi-acoustic and optical modes scatter more severely compared to the low-frequency, high group velocity modes. As indicated in Ref. [47, 50], the stronger influence of the boundary scattering originates from the mismatch in the flat dispersion modes along the transport direction of the NW caused by the rough geometry. Note that the stronger boundary scattering for higher frequencies is even the case for different materials such as graphene nanoribbons as we showed in Ref. [46] through non-equilibrium Green's function simulations. Therefore, under weak roughness conditions, the thermal conductivity could still show an increase, or at least a weaker decrease as the diameter is reduced. Under stronger roughness conditions in ultra-narrow NWs, i.e. roughness amplitude as large as 20% of the diameter, however, the phonon mean free path and the thermal conductivity are very close to what the Casimir limit predicts (fully diffusive phonons at all frequencies, with phonon MFPs limited by the NW diameter) as discussed in Ref. [45], and the effects we describe might not be noticeable.

## IV. Conclusions

In this work we study the thermal properties of ultra-thin silicon nanowires using the atomistic modified valence-force-field method for the computation of the phonon bandstructure and the Landauer ballistic transport as well as the diffusive Boltzmann transport methods. We show that the *"problem of long-wavelength phonons"* as described by Ziman and others, which causes divergence in the thermal conductivity of quasi-1D channels with increasing length, is also present in Si NWs. The divergence occurs not only as the length is increased, but as the diameter is reduced as well. We connect this to the fact that in ultra-narrow nanowires the density-of-states and the transmission function of long-wavelength phonons acquires a finite value, as compared to zero in the bulk materials, which increases their importance in carrying heat, and results in thermal transport improvement as the diameter is reduced below 5nm. This results in a striking anomalous increase in the thermal conductance (at low temperatures) and thermal conductivity (at all temperatures we examine – up to T=300K) as the diameter is reduced below 5nm. The consequence of this effect is that a larger portion of heat is carried by low frequency phonons in ultra-narrow nanowires as compared to bulk, i.e. almost 80% of the heat is carried by phonons with energies below 5meV and predominantly long mean-free-paths.

## Acknowledgements

The work leading to these results has received funding from the European Community's Seventh Framework Programme under grant agreement no. FP7-263306.

Figure 1:

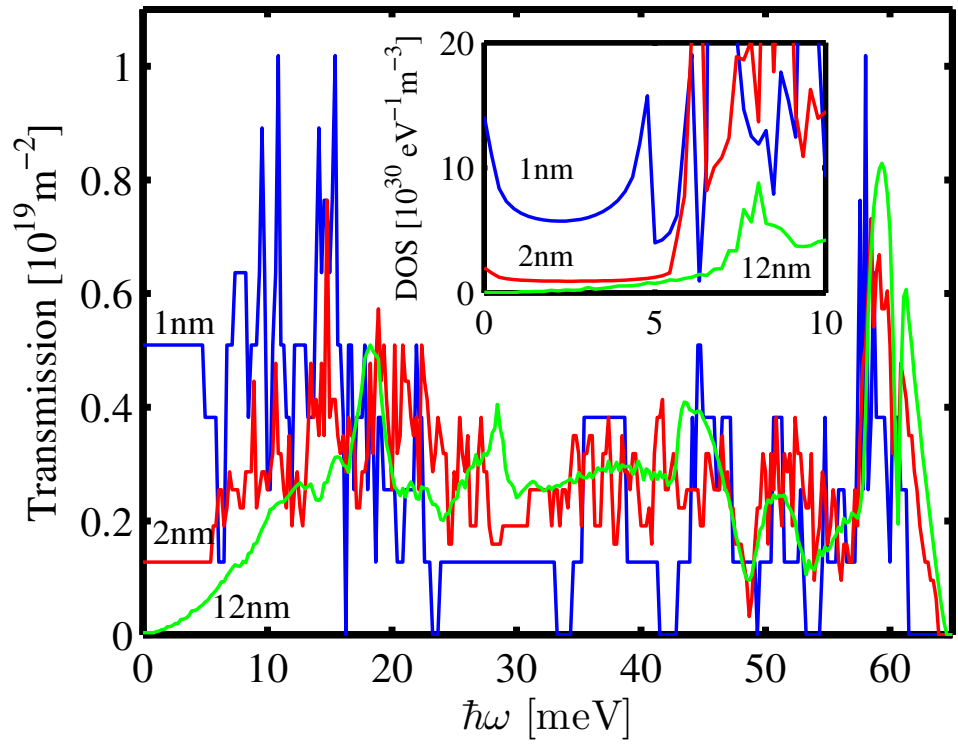

Figure 1 caption:

The normalized transmission function (number of propagating modes per unit area) for Si nanowires of diameters $D$=1nm (blue), 2nm (red) and 12nm (green). Inset: The density of phonon states per unit volume and energy for the same NW diameters.

Figure 2:

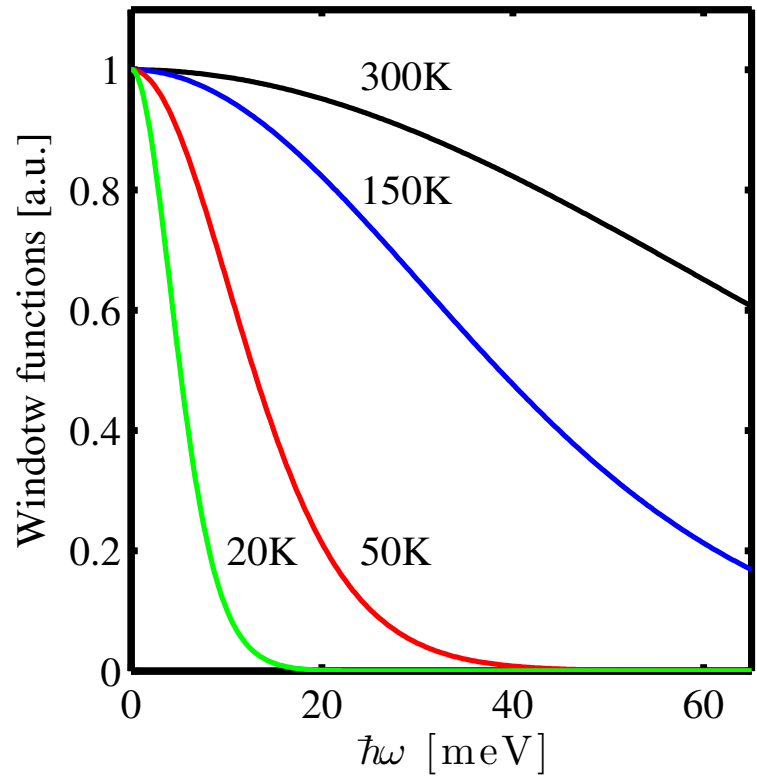

Figure 2 caption:

The phonon window function that determines the thermal conductance in the case of ballistic transport conditions for various temperatures. As the temperature decreases, the window function becomes narrower.

Figure 3:

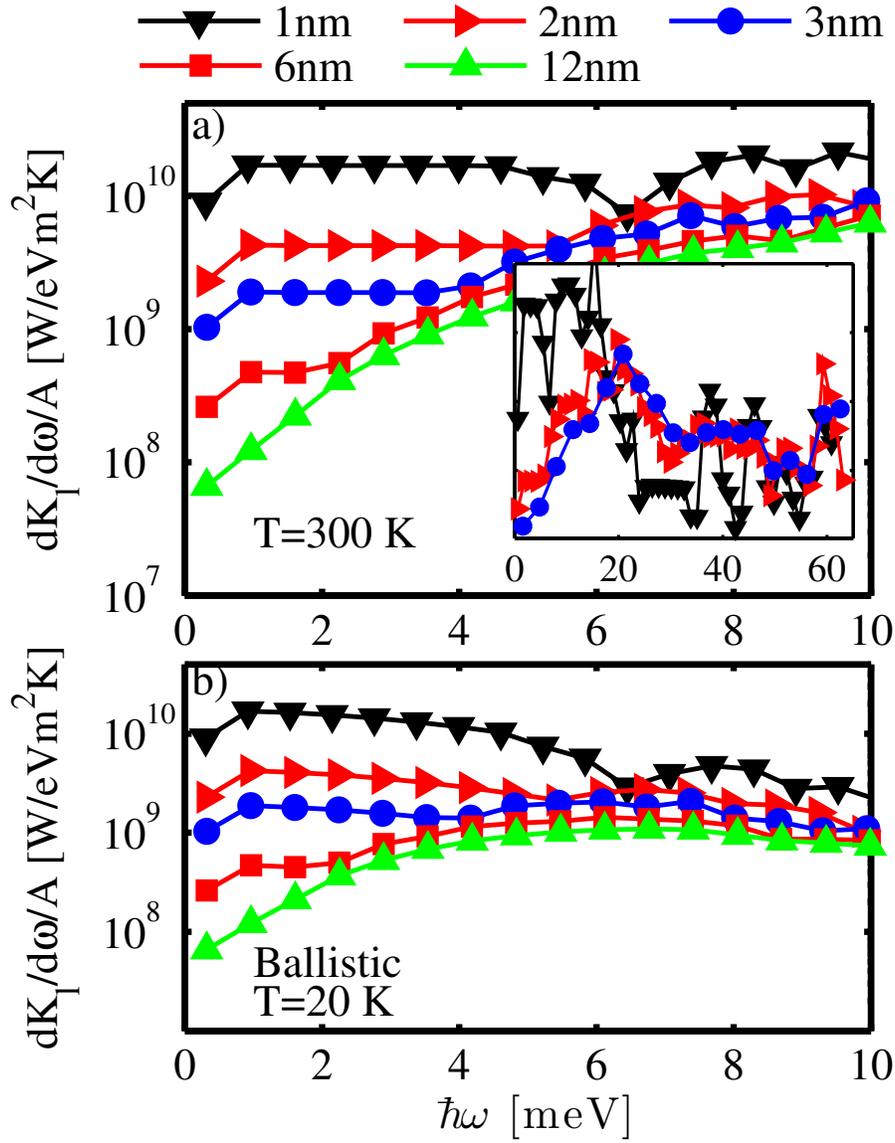

Figure 3 caption:

(a) The differential contribution to the ballistic thermal conductance of low frequency phonons of energies up to 10meV at room temperature T=300K. Nanowire diameters $D$=1nm (black-triangle), $D$=2nm (red-triangle), $D$=3nm (blue-circle), $D$=6nm (red-square), and $D$=12nm (green-triangle) are shown. The contribution of the low frequency modes increases with decreasing diameter. Inset: The differential contribution to the ballistic thermal conductance in the entire energy spectrum. (b) The same as in (a) for low temperatures T=20K.

Figure 4:

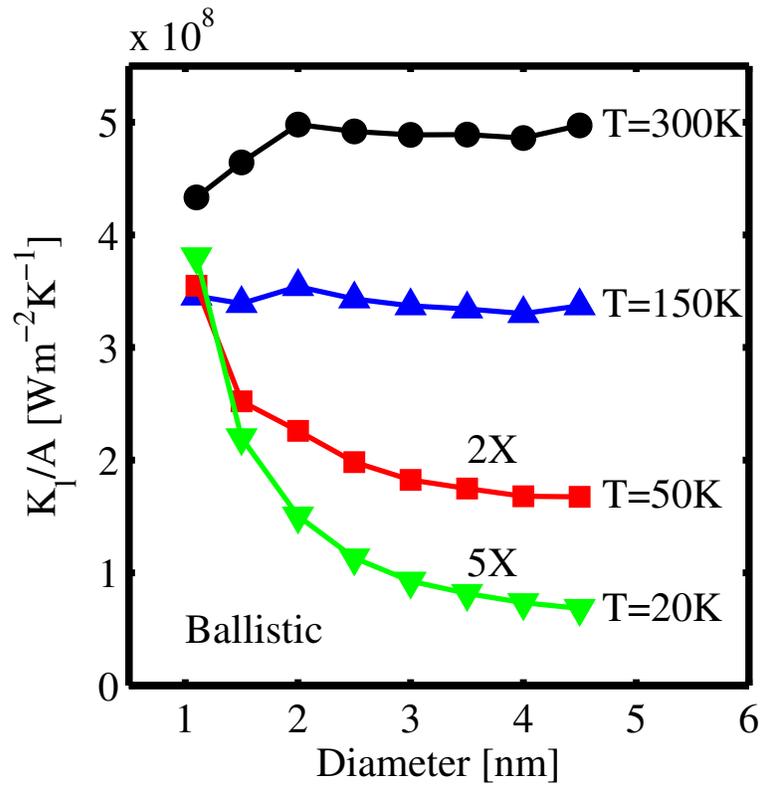

Figure 4 caption:

Ballistic thermal conductance of Si NWs versus diameter for different temperatures. The increase in conductance with diameter scaling is noted when applicable.

Figure 5:

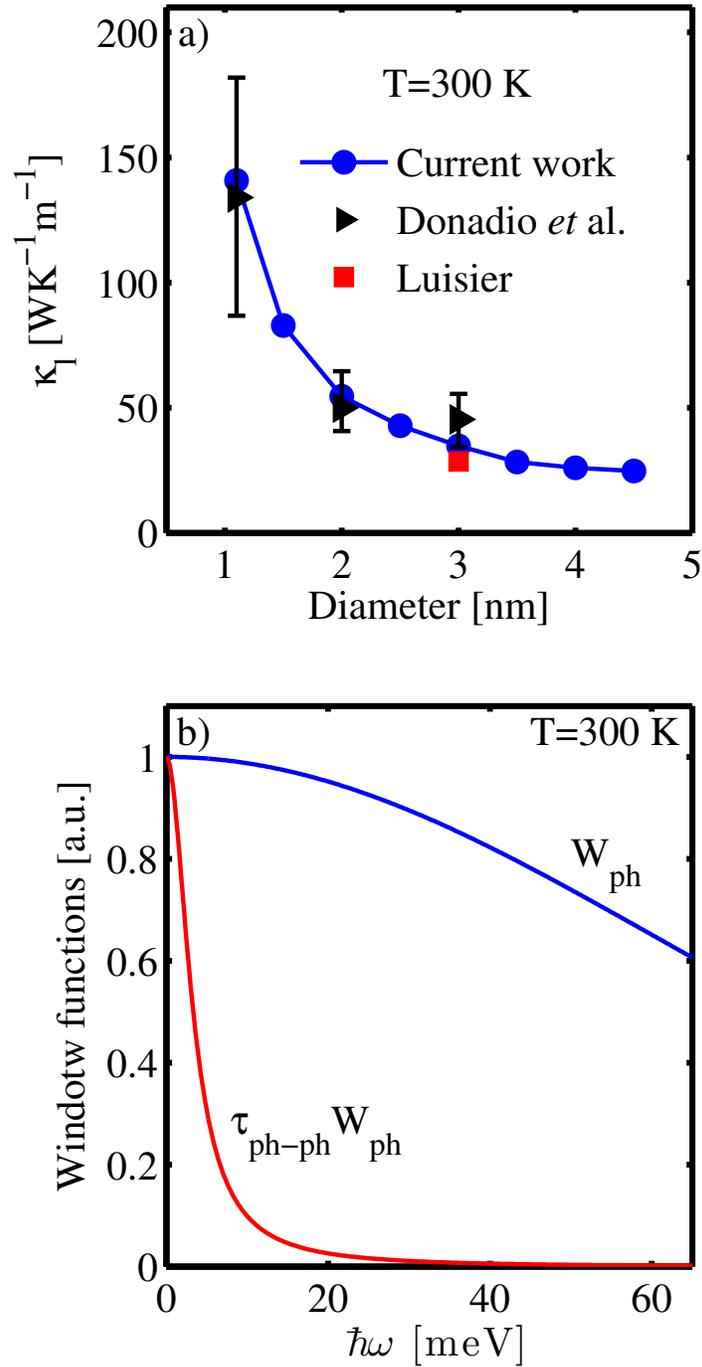

Figure 5 caption:

(a)Room temperature phonon-phonon limited thermal conductivity for Si nanowires versus diameter. Blue-dots: This work. Black-triangle: Molecular dynamics results by Donadio *et* al. [10]. Red-square: Calculations by Luisier [40]. (b) The phonon window function that determines the thermal conductance in the case of ballistic transport denoted $W_{ph.}$ (blue line) and phonon-phonon scattering diffusive transport denoted $\tau_{ph.-ph.}W_{ph.}$ (red line). With the introduction of scattering the window function becomes narrower.

Figure 6:

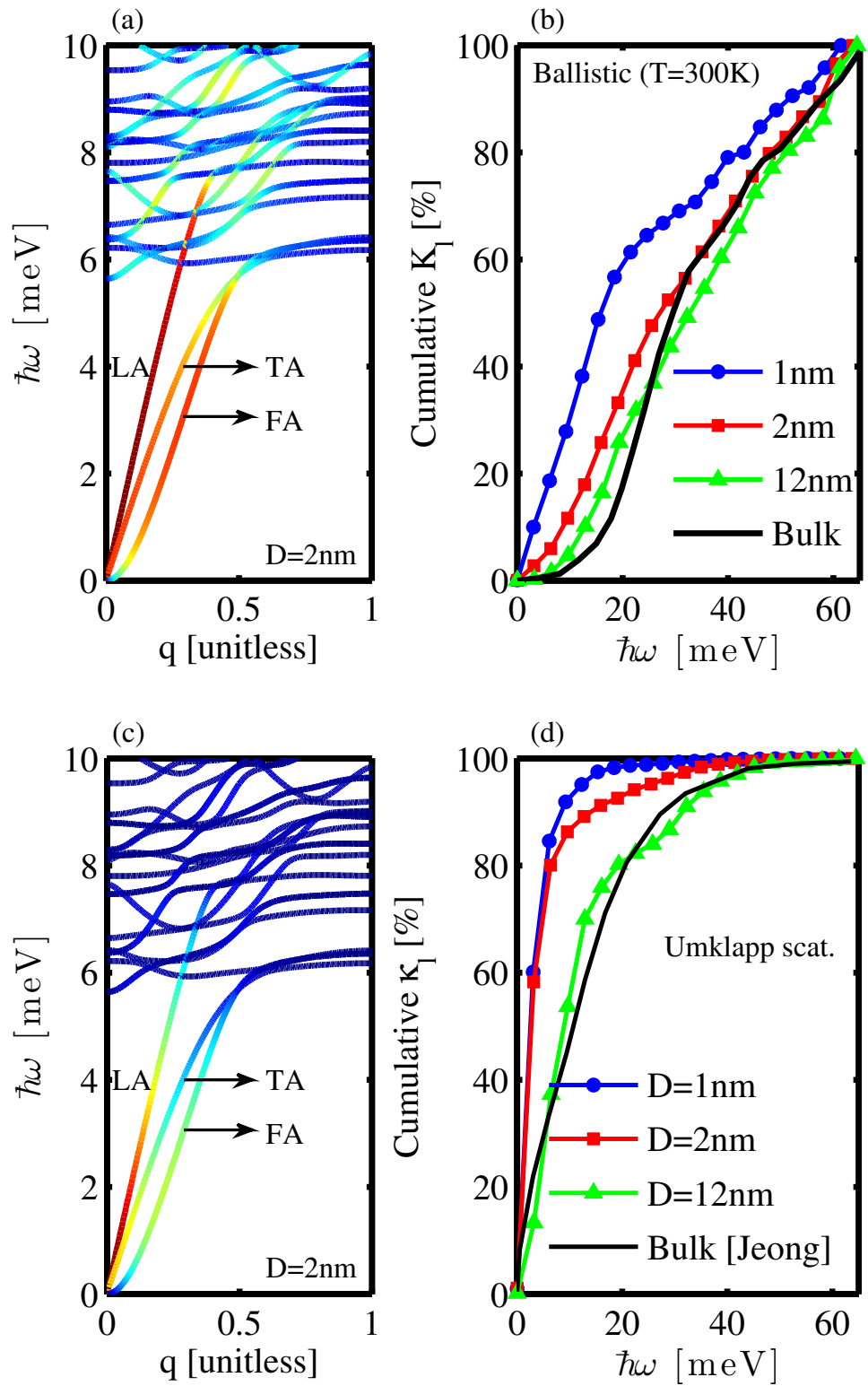

## Figure 6 caption:

(a) The phonon dispersion of the $D$=2nm <100> nanowire. The colorplot shows the contribution of various phonon states to the ballistic thermal conductance (red indicates the highest, and blue the lowest thermal conductance). (b) The cumulative ballistic thermal conductance versus energy for NWs of $D$=1nm (blue-dots), 2nm (red-square), and 12nm (green-triangle). The black-solid line shows the cumulative thermal conductance of bulk Si from Jeong *et* al. [39]. (c-d) Same as in (a-b) but phonon-phonon scattering conditions are assumed in the calculation of the thermal conductivity.

Figure 7:

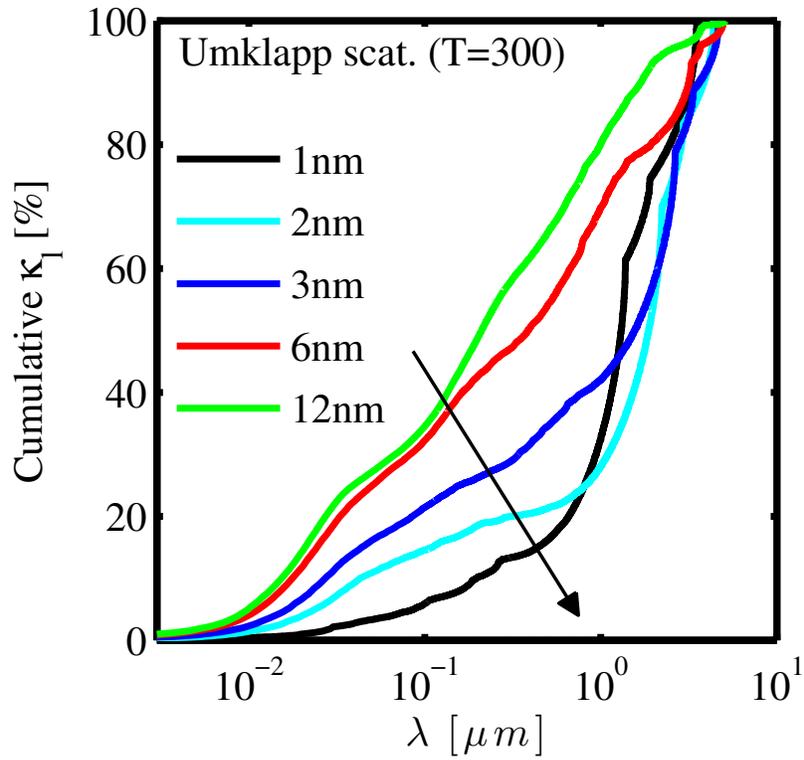

Figure 7 caption:

The cumulative thermal conductivity versus phonon mean-free-path for NWs with diameters $D$=12nm down to $D$=1nm at room temperature under phonon-phonon limited scattering conditions.